\documentclass[12pt]{article}
\usepackage[utf8]{inputenc}
\usepackage{url} 
\usepackage[
    backend=biber,
    style=apa,
    uniquelist=false
  ]{biblatex}
\usepackage{dirtytalk}
\usepackage{amsthm}
\usepackage{amssymb}
\usepackage{amsmath}
\usepackage{mathtools}
\usepackage{xfrac}
\usepackage{gensymb}
\usepackage{graphicx}
\usepackage{multirow}
\usepackage{setspace}
\usepackage{caption}
\usepackage{float}
\usepackage{makecell}
\newcommand{\defeq}{\vcentcolon=}
\usepackage{comment}
\usepackage{ifthen}
\usepackage[papersize={8.5in,11in}, margin=1in]{geometry}
\usepackage{sectsty}
\subsectionfont{\itshape}

\title{Question-driven ensembles of flexible ETAS models}
\author{Leila Mizrahi$^{1, *}$, Shyam Nandan$^1$, William Savran$^2$, \\ Stefan Wiemer$^1$, Yehuda Ben-Zion$^2$}
\usepackage{nameref}
\begin{document}

\maketitle

$^1$ Swiss Seismological Service, ETH Zurich, Switzerland\\

$^2$ University of Southern California, Los Angeles CA, United States\\

$^*$ corresponding author: leila.mizrahi@sed.ethz.ch, Sonneggstrasse 5, 8092 Zurich, Switzerland\\

The authors acknowledge there are no conflicts of interest recorded.
\newpage

\begin{abstract}
    
    The development of new earthquake forecasting models is often motivated by one of the following complementary goals: to gain new insights into the governing physics and to produce improved forecasts quantified by objective metrics.
    Often, one comes at the cost of the other.
    Here, we propose a question-driven ensemble (QDE) modeling approach to address both goals.
    We first describe flexible ETAS models in which we relax the assumptions of parametrically defined aftershock productivity and background earthquake rates during model calibration. 
    Instead, both productivity and background rates are calibrated with data such that their variability is optimally represented by the model.
    Then we consider 64 QDE models in pseudo-prospective forecasting experiments for Southern California and Italy.
    QDE models are constructed by combining model parameters of different ingredient models, where the rules for how to combine parameters are defined by questions about the future seismicity.
    The QDE models can be interpreted as models which address different questions with different ingredient models.
    We find that certain models best address the same issues in both regions, and that QDE models can substantially outperform the standard ETAS and all ingredient models.
    The best performing QDE model is obtained through the combination of models allowing flexible background seismicity and flexible aftershock productivity, respectively, where the former parameterizes the spatial distribution of background earthquakes and the partitioning of seismicity into background events and aftershocks, and the latter is used to parameterize the spatio-temporal occurrence of aftershocks.

\end{abstract}
\section*{Introduction}
    
    Earthquake forecasting is one of the defining problems of seismology. To provide useful solutions, forecasting models use a wide range of approaches:
    Coulomb rate-and-state (CRS) models (\cite{cocco2010sensitivity}; \cite{parsons2012evaluation}; \cite{mancini2019improving}) calculate Coulomb stress changes and couple them with a lab-based constitutive friction law (\cite{dieterich1994constitutive}).
    On the other end of the spectrum are statistical models, with the Epidemic-Type Aftershock Sequence (ETAS) model being the best performing current statistical approach (\cite{cattania2018forecasting}; \cite{taroni2018prospective}).
    First introduced by \textcite{ogata1988statistical}, it models seismicity rate as the sum of background and aftershock events, where aftershocks are triggered according to regional empirical laws.
    In-between the purely physics-based and purely statistics-based approaches are models such as the short-term earthquake probability (STEP) model (\cite{gerstenberger2005real}), the Inlabru model (\cite{bayliss2020data}) and hybrid Coulomb/statistical models (\cite{steacy2014new}). The STEP model combines clustering principles with fault information in a statistical model to produce time-dependent forecasts. The Inlabru model more generally allows the inclusion of diverse data sets as covariates to issue time-independent seismicity forecasts. A hybrid Coulomb/statistical model redistributes seismicity forecasted by STEP according to Coulomb stress changes.
    
    While physics-based models aim to describe the processes and mechanisms underlying seismogenesis, statistical models are generally more empirical and  data-driven.
    Ultimately, ``all models are wrong, but some are useful", to cite the famous statistician George \textcite{box1979robustness}.
    Usefulness can be viewed from different perspectives. Different forecasting models can be useful for gaining new scientific insight, for producing the most accurate forecasts, or for producing forecasts that are most suited for operational earthquake forecasting (OEF), given the trade-off between accuracy and computational cost.
    \textcite{cattania2018forecasting} found in a pseudo-prospective forecasting experiment for the 2010-2012 Canterbury, New Zealand earthquake sequence that hybrid Coulomb/statistical models have a similar forecasting skill as CRS models, at a lower computational effort.
    \textcite{mancini2019improving} and \textcite{mancini2020predictive} conducted pseudo-prospective experiments for the 2016 central Italy and the 2019 Ridgecrest, California sequences, comparing CRS models of different complexity with ETAS forecasts.
    In both studies, the forecasting skill of CRS models increases with their complexity, with the most complex CRS model performing similarly to ETAS.
    \textcite{hardebeck2021spatial} investigated possible reasons for the general underperformance of the physics-based models relative to statistical models and suggested that understanding and incorporating heterogeneities in background conditions into physical forecasting models may be key to improving their skill.
    
    Having been tested thoroughly and systematically (\cite{woessner2011retrospective}; \cite{ogata2013comprehensive}; \cite{strader2017prospective}; \cite{taroni2018prospective}; \cite{nandan2019forecastingrates}; \cite{savran2020pseudoprospective}), ETAS models meanwhile remain the state-of-the art of earthquake forecasting and are being used or considered for OEF at various locations (\cite{marzocchi2014establishment}; \cite{rhoades2016retrospective}; \cite{field2017synoptic}; \cite{nandan2021global}; \cite{kamer2021democratizing}; \cite{van2022prospective};).
    Besides using the most basic formulation of ETAS, modelers also commonly refine the model.
    For instance, \textcite{bach2012improving} enhanced ETAS with fault information, ShakeMaps, ground motion models, or Coulomb stress changes.
    \textcite{seif2017estimating} assessed the biasing effects of data incompleteness and model assumptions on the estimated ETAS parameters.
    Several techniques have been proposed to address the effects of short-term aftershock incompleteness (\cite{mizrahi2021effect}; \cite{hainzl2022etas}; \cite{grimm2022solving}) or the assumption of isotropic aftershock triggering (\cite{grimm2022solving}; \cite{page2022aftershocks}).
    Other studies focus on deriving spatial variations of ETAS parameters or background seismicity (\cite{nandan2017objective}; \cite{nandan2021seismicity}; \cite{enescu2009correlations}), also relating parameter variations with physical quantities such as heat flow.
    Others have refined the standard ETAS model with a relationship between magnitudes of triggered and triggering earthquakes and a magnitude-dependent Omori kernel and found the resulting models to possess improved forecasting performance (\cite{nandan2021global}; \cite{nandan2019magnitude}). A recent framework for modeling seismicity with an invariant Galton–Watson stochastic branching process provides a generalization of ETAS that is invariant with respect to various common deficiencies of earthquake catalogs (\cite{kovchegov2022invariant}). However, this framework has not yet been used for forecasting seismicity.
    
    A related forecasting topic which has recently received attention is ensemble modeling (\cite{rhoades2009mixture}; \cite{marzocchi2012bayesian}; \cite{taroni2014assessing}; \cite{bird2015gear1}; \cite{akinci2018ensemble}; \cite{llenos2019ensembles}; \cite{bayona2021two}).
    The idea, widely used for decades in the meteorological and climate forecasting community (\cite{tracton1993operational}; \cite{leutbecher2008ensemble}; \cite{eyring2016overview}), is to combine different models in an overarching ensemble model to obtain more robust forecasts.
    Commonly, an ensemble is a linear or multiplicative combination of ingredient models (e.g. \cite{bird2015gear1}), and the challenge is to optimize the weights given to each model.
    In a recent study, \textcite{bayona2021two} found that the time-independent ensemble models WHEEL and GREAR1 (\cite{bird2015gear1}) outperform the ingredient models of which they consist.
    \textcite{akinci2018ensemble} found that their time-independent ensemble model outperforms its ingredients and performs similarly to the best-performing time-independent model tested in the 2009 CSEP experiment (\cite{zechar2010collaboratory}; \cite{schorlemmer2010first}) for Italy. 
    In the context of time-dependent models, \textcite{taroni2014assessing} and  \textcite{gerstenberger2014seismic} used ensemble approaches, and \textcite{llenos2019ensembles} found that ensembles of ETAS models perform best for the 2015 San Ramon, California Swarm.
    \textcite{shebalin2014combining} proposed an iterative method to combine forecasting models and found the resulting models to have advantageous properties compared to the ingredient models or traditional linear combinations thereof.
    The emerging consensus across the mentioned studies is that ensemble modeling is a promising path to use for earthquake forecasting; this is also demonstrated by the fact that they are currently implemented in Italy's OEF system (\cite{marzocchi2014establishment}).
    Yet, a breakthrough of ensemble models as established in the meteorological forecasting community is still pending.\\

    For practical operational forecasting, especially in regions that are less studied due to a lack of data or resources, a balance must be achieved between model accuracy and simplicity.
    With this in mind, we relax some of the assumptions behind ETAS.
    We allow aftershock productivity and background seismicity to be described non-parametrically, providing event-specific productivity and background rates.
    This aims to better capture the real behavior of seismicity without making any choices on resolution, parametric form, etc.
    Using pseudo-prospective forecasting experiments in Southern California and Italy, we evaluate whether these flexible ETAS (flETAS) models provide superior forecasts.
    
    We also propose a novel approach for question-driven ensemble (QDE) modeling, fundamentally different from traditional ensemble modeling approaches.
    In the QDE approach, models are combined in the parameter space as opposed to the solution space.
    Several ETAS-like models are fit to the observed data, yielding an individual set of parameters for each model.
    A QDE model is then created by defining a new set of parameters based on a combination of the ingredient model parameters.
    The rules to combine parameters are defined by dividing the forecasting problem into several sub-problems. 
    Each sub-problem addresses a question regarding the number of forecasted events or the spatio-temporal distribution of either background earthquakes or aftershocks.
    A QDE model can be viewed as a model which addresses different questions with different ingredient models. 
    This approach allows the combination of ETAS variants but can be extended to combining more general types of seismicity models.
    
    By including such QDE models in the forecasting experiments, we assess their forecasting capability in comparison with their ingredient models, standard ETAS and flETAS.
    At the same time, the QDE approach helps to understand which ingredient models are best suited to solve different forecasting sub-problems, thus, making it useful from the perspective of gaining new scientific insight.\\

    The remainder of this paper is structured as follows.
    We describe flETAS models and the QDE approach in the next section \nameref{sec:model}.
    The setup for the forecasting experiments, the data analyzed and the metrics used to evaluate forecasting performance are described in section \nameref{sec:forecast}.
    We present and discuss our results in \nameref{sec:results} and finally provide our \nameref{sec:concl}.

\section*{Flexible ETAS Models}
    \label{sec:model}
    
    The following sub-sections describe flexible ETAS models and explain the question-driven ensemble modeling. We begin by explaining the algorithm used to estimate the parameters of the ETAS model. Then, we describe how to relax some parametric assumptions of the ETAS model. Finally, we introduce a framework for question-driven ensemble modeling of flexible ETAS models.
    
    \subsection*{Expectation Maximization Algorithm}
        Consider an earthquake catalog
        
        \begin{equation}
            C=\{e_i=(m_i, t_i, x_i, y_i), i\in\{1, \dots, n\}\}
        \end{equation}
        consisting of events $e_i$ of magnitudes $m_i$ which occur at times $t_i$ and locations $(x_i, y_i)$.
        
        The ETAS model describes earthquake rate as
        
        \begin{equation}
        \label{eq:etasrate}
            \lambda(t,x,y | \mathcal{H}_t) = \mu + \sum_{i: t_i < t} g(m_i, t-t_i, x-x_i, y-y_i).
        \end{equation}
        That is, the sum of background rate $\mu$ and the rate of all aftershocks of previous events $e_i$.
        The aftershock triggering rate $g(m, \Delta t, \Delta x, \Delta y)$ describes the rate of aftershocks triggered by an event of magnitude $m$, at a time delay of $\Delta t$ and a spatial distance $(\Delta x, \Delta y)$ from the triggering event.
        We use here the definition
        
        \begin{equation}
            \label{eq:g_etas}
            g(m,\Delta t, \Delta x,\Delta y) = \frac{k_0 \cdot e^{a(m-m_{ref})}\cdot e^{-\Delta t/ \tau}}{\left((\Delta x^2 + \Delta y^2) + d\cdot e^{\gamma (m-m_{ref})}\right)^{1+\rho} \cdot (\Delta t+c)^{1+\omega}},
        \end{equation}
        as in \textcite{nandan2021global} and \textcite{mizrahi2021embracing}.
        This formulation differs from other, more commonly used formulations of ETAS models in that it uses an Exponentially Tapered Omori Kernel (ETOK).
        In their paper, \textcite{nandan2021global} compare the ETAS model with ETOK to a more general version thereof (MDOK) which allows a magnitude dependency, finding that the more general version allows better forecasts. 
        This indicates that including an exponential taper does lead to improved forecasts when compared to the commonly used Omori kernel.
        Besides allowing less heavy tails in the temporal distribution of aftershocks, this formulation of the Omori kernel makes it possible for the parameter $\omega$ to attain negative values, which is not possible in the traditional formulation.
        Note also that our choice of this base model does not impact the main conclusions that can be drawn from comparing it to modified versions of itself.  

        \paragraph{}
        To calibrate the ETAS model, the nine parameters to be optimized are the background rate $\mu$ and $k_0, a, c, \omega, \tau, d, \gamma, \rho$, which parameterize the aftershock triggering rate $g(m, t, x, y)$ given in Equation \ref{eq:g_etas}. Implicitly, the model assumes that only earthquakes with magnitudes larger than or equal to $m_{ref}$ can trigger aftershocks. Most applications of the method define $m_{ref}$ as equal to the constant value of $m_c$.
        
        We build on the expectation maximization (EM) algorithm to estimate the ETAS parameters (\cite{veen2008estimation}). In this algorithm, the expected number of background events $\hat n$ and the expected number of directly triggered aftershocks $\hat l_i$ of each event $e_i$ are estimated in the expectation step (E step), along with the probabilities $p_{ij}$ that event $e_j$ was triggered by event $e_i$, and the probability $p_{j}^{ind}$ that event $e_j$ is independent. Following the E step, the nine parameters are optimized to maximize the complete data log-likelihood in the maximization step (M step). 
        E and M steps are repeated until convergence of the parameters.
        The usual formulation of the EM algorithm defines
        
        \begin{align}
            \hat n &= \sum_j p_j^{ind},\\
            \hat l_i &= \sum_j p_{ij}\label{eq:lhat},
        \end{align}
        
        and
        
        \begin{align}
            p_{ij} &= \frac{g_{ij}}{\mu + \sum_{k: t_k <t_j} g_{kj}},\label{eq:pij}\\
            p_{j}^{ind} &= \frac{\mu}{\mu + \sum_{k: t_k <t_j} g_{kj}} \label{eq:p_ind},
        \end{align}
        with $g_{kj} = g(m_k, t_j-t_k, x_j-x_k, y_j-y_k)$ being the aftershock triggering rate of $e_k$ at location and time of event $e_j$. 
        For a given target event $e_j$, Equations (\ref{eq:pij}-\ref{eq:p_ind}) define $p_{ij}$ to be proportional to the aftershock occurrence rate $g_{ij}$, and $p_j^{ind}$ to be proportional to the background rate $\mu$.
        As an event must be either independent or triggered by a previous event, the normalization factor 
        $\Lambda_j \defeq \mu + \sum_{k: t_k <t_j} g_{kj}$
        in the denominator of Equations \ref{eq:pij}-\ref{eq:p_ind} stipulates that $p_j^{ind} + \sum_{k: t_k <t_j} p_{kj} = 1$.

    \subsection*{Introducing Flexibility}
        
        In the above formulation of the ETAS model, the the rate of background earthquakes is described by the parameter $\mu$, which does not vary with space nor time.
        During the maximization step of the EM algorithm, $\mu$ can be estimated independently from the other parameters as 
        
        \begin{align}
            \mu = \frac{\hat n}{A_{R} \cdot T},
        \end{align}
        
        where $A_R$ and $T$ denote the area of the study region and the length of the considered time window, respectively. In some
        approaches, the region of interest is divided into several sub-regions which can have their own values for $\mu$ (\cite{veen2008estimation}).
        An iterative algorithm to estimate spatial variations of background rate based on maximum likelihood estimation used 
        a Gaussian kernel smoothing (\cite{zhuang2012long}) to the catalog event locations, weighted by their estimated independence probability, to obtain an estimate of $\mu(x, y)$.
        Here, we present a similar approach using expectation maximization, which has been shown to be more stable with respect to the initial conditions compared to maximum likelihood approaches (\cite{veen2008estimation}).
        Our approach is similar yet not identical to the one described by \textcite{nandan2021seismicity} which uses a regularized inverse power law for smoothing the locations.
        We define the background rate at a location $(x, y)$ as
        
        \begin{align}
            \label{eq:mu_fletas}
            \mu(x, y) = \frac{1}{T} \cdot \sum_j p_j^{ind} \cdot k(\Delta x_j, \Delta y_j),
        \end{align}
        
        where $k(\Delta x_j, \Delta y_j)$ is the Gaussian kernel with bandwidth $\sigma$ applied to the distance $(\Delta x_j, \Delta y_j)$ of event $e_j$ to the location $(x, y)$,
        
        \begin{align}
            k(\Delta x, \Delta y) = \frac{1}{2\pi \sigma^2} \cdot \exp({-\tfrac{1}{2}\cdot \tfrac{\Delta x^2 + \Delta y^2}{\sigma^2}}).
        \end{align}
        
        The bandwidth $\sigma$ determines the smoothness of the background event density.
        In principle, $\sigma$ could be calibrated itself, but we choose to fix it to $5km$ for simplicity.
        Our next modification to the standard ETAS model is to allow flexibility of the aftershock probability. 
        The number of directly triggered aftershocks $\hat l_j$ is estimated during the expectation step of the EM algorithm as described in Equation (\ref{eq:lhat}).
        We can thus replace the term $k_0\cdot e^{a(m-m_{ref})}$ in Equation (\ref{eq:g_etas}) with $\kappa_j$, where $\kappa_j$ is stipulated to be proportional to $\hat l_j$.
        Instead of parameterizing aftershock productivity to be exponentially increasing with the magnitude of the triggering event, we allow each event to have its own productivity.
        This yields
        
        \begin{align}
            \label{eq:g_etas_fp}
            g_{j_{\theta, \kappa_j}}(m,\Delta t, \Delta x,\Delta y) = \frac{\kappa_j \cdot e^{\Delta t/ \tau}}{\left((\Delta x^2 + \Delta y^2) + d\cdot e^{\gamma (m-m_{ref})}\right)^{1+\rho} \cdot (\Delta t+c)^{1+\omega}}
        \end{align}
        
        for given parameters $\theta = (c, \omega, \tau, d, \gamma, \rho)$ and $\kappa_j$. 
        The EM algorithm is adapted as follows:
        
        \begin{enumerate}
            \item Define initial estimates of $\kappa_j$ as $\kappa_j = e^{a(m_j - m_{ref})}$ with a random guess for $a$.
            \item Define initial estimates of independence probability $p_j^{ind} \equiv 0.1$. The inversion result is not sensitive to this choice.
            \item Define random initial guesses for the parameters $\theta = (c, \omega, \tau, d, \gamma, \rho$).
            \item \label{Estep} Expectation Step: Calculate $\hat n, \hat l_j, p_{ij}, p_j^{ind}$ using the current estimates of $\kappa_j, \theta,$ and $p_j^{ind}$. $p_{ij}, p_j^{ind}$ are calculated using Equations (\ref{eq:pij}-\ref{eq:p_ind}), but using the flexible definitions of $g_{ij}$ and $\mu(x, y)$ of Equations (\ref{eq:mu_fletas}) and (\ref{eq:g_etas_fp}).
            \item Maximization Step: Optimize the parameters $\theta$ to minimize the complete data log likelihood (see \textcite{mizrahi2021embracing} for details), given the current estimates of $\hat n, \hat l_j, p_{ij}, p_j^{ind}$. 
            \item Update $\kappa_j^{new}$ to be $\kappa_j^{old} \cdot \tfrac{\hat l_j}{G_{j_{\theta, \kappa_j^{old}}}},$ where $G_{j_{\theta, \kappa_j^{old}}}$ is the expected total number of aftershocks of $e_j$, given $\theta$ and $\kappa_j^{old}$.
            This ensures that $\hat l_j = G_{j_{\theta, \kappa_j^{new}}}$.
            We calculate $G_{j_{\theta, \kappa_j}}$ as
            
            \begin{align}
                G_{j_{\theta, \kappa_j}} = \iint_R\int_0^{t_{end}-t_j} g_{j_{\theta, \kappa_j}}(m_j, t, x, y) \,dt\,dx\,dy,
            \end{align}
            where $t_{end}$ is the end time of the considered time window, and we assume the spatial region $R$ to extend infinitely in space, allowing a facilitated, asymptotically unbiased estimation of ETAS parameters (\cite{schoenberg2013facilitated}).

            \item Repeat from \ref{Estep} until convergence of $\theta$, i.e. until $\sum_{a_i \in \theta} |a_i^{new}-a_i^{old}| < 10^{-3}$.
        \end{enumerate}
      
        After the inversion, we calibrate an overall productivity law for the flETAS models with free productivity to avoid over-fitting with event-wise productivity.
        From the individually estimated productivities $\kappa_j$ of magnitude $m_j$ events, we calibrate a law of the form 
            
            \begin{equation}
                \label{eq:kappa_simple}
                \kappa(m) = k_0 \cdot e^{a(m-m_{ref})}
            \end{equation}

        by minimizing the sum of absolute residuals between the observed $\bar \kappa (m) = \frac{1}{n(m)}\sum_{i:m_i = m} \kappa_i$ and the theoretical $\kappa(m) = k_0\cdot e^{a(m-m_{ref})}$, where $n(m)$ is the number of events with magnitude $m$.

        Then, productivity is treated the same way as in the case of standard ETAS. 
        In this way, the variability of productivity is only accounted for during the parameter inversion process and may lead to more accurate estimators of the productivity as well as the remaining ETAS parameters.

    \subsection*{Question-driven ensemble (QDE) modeling}
    
        We propose a novel approach for question-driven ensemble (QDE) modelling, where a forecast is created by combining model parameters of different ingredient models. 
        The rules for how parameters can be combined are defined by questions which divide the forecasting problem into several sub-problems: \textit{How many background events are expected? Where are they expected ? When are they expected? How many aftershocks are expected? Where are they expected? When are they expected?}
        
        By answering each of these questions with different ingredient models, we create a suite of ensembles.
        The remainder of this section establishes rules to combine parameters based on the questions.\\
        
        Consider a collection of ETAS or flETAS ingredient models, $(M_{i})_{i=0, \dots, n_M}$.
        As they are sufficiently defined through their parameters, we can write 
        
        \begin{equation}
            M_i = (\mu_i, \kappa_i, c_i, \omega_i, \tau_i, d_i, \gamma_i, \rho_i).
        \end{equation}
        
        In case $M_i$ is a flETAS model, $\mu_i = \mu_i(x, y)$ can vary with space.
        For simplicity, we denote with $\kappa_i$ the function which assigns to each event its appropriate value to replace the term $\kappa_j$ in Equation (\ref{eq:g_etas_fp}).
        In our case, this means that we define $\kappa_i(m)= k_{0_i} \cdot e^{a_i(m-m_{ref})}$, where $k_{0_i}$ and $a_i$ are either obtained during parameter inversion directly, or afterwards in case $M_i$ is a flETAS model with free productivity.
        We chose the notation of $\kappa_i$ instead of ($k_{0_i}, a_i$) to emphasize this possible distinction.
        We can then generally describe the aftershock triggering kernel $g$ as
        
        \begin{align}
            \label{eq:g_fletas}
            g_i(m,\Delta t, \Delta x,\Delta y) = \frac{\kappa_i \cdot e^{\Delta t/ \tau_i}}{\left((\Delta x^2 + \Delta y^2) + d_i\cdot e^{\gamma_i (m-m_{ref})}\right)^{1+\rho_i} \cdot (\Delta t+c_i)^{1+\omega_i}}.
        \end{align}
        \\
        
        Let us now revisit the questions above.
        
        \begin{enumerate}
            \item \textit{How many background events are expected?}\\
            \label{qu:nb}
            More precisely, what we want to ask here is how many background events do we expect in total in the region $R$ and forecasting horizon $[T_0, T_1]$ we are issuing a forecast for.
            The answer to this question, given out of the perspective of model $M_i$, is
            
            \begin{equation}
                N_{B_i} = \iint_R\int_{T_0}^{T_1} \mu_i(x, y) \,dt\,dx\,dy.
            \end{equation}
            
            \item \textit{Where and when are they expected?}\\
            \label{qu:db}
            We address for now these two questions jointly. 
            The spatio-temporal density of background events is given by
            
            \begin{equation}
                f_{B_i}(x, y, t) = \frac{\mu_i(x, y)}{\iint_R\int_{T_0}^{T_1} \mu_i(x, y) \,dt\,dx\,dy} = \frac{\mu_i(x, y)}{N_{B_i}},
            \end{equation}
            which is effectively time-independent due to our choice of a time-independent $\mu(x, y)$.
            
            \item \textit{How many aftershocks are expected?}\\
            \label{qu:na}
            Again, what we want to ask here is how many aftershocks do we expect in total in the region $R$ and forecasting horizon $[T_0, T_1]$ we are issuing a forecast for.
            For an individual event $e_j$, we expect it to have $n_A$ aftershocks, where
            
            \begin{equation}
                n_{A_i}(e_j) = \iint_R\int_{T_0}^{T_1} g_i(m_j, t-t_j, x-x_j, y-y_j)\,dt\,dx\,dy.
            \end{equation}
            
            The total number of aftershocks $N_{A_i}$ is then given as the sum of aftershocks of all events
            \begin{equation}
                N_{A_i} = \sum_{j: t_j < T_1} n_{A_i}(e_j).
            \end{equation}
            
            \item \textit{Where and when are they expected?}\\
            \label{qu:da}
            We again answer these two questions jointly. 
            If we define
            
            \begin{equation}
                G_i(x, y, t) \defeq \sum_{j: t_j < T_1} g_i(m_j, t-t_j, x-x_j, y-y_j)
            \end{equation}
    
            as the total rate of aftershocks at time $t$ and location $(x, y)$, consisting of the sum of aftershock rates of all events that occurred prior to the end $T_1$ of the forecasting horizon, the spatio-temporal density of aftershocks is given by
            
            \begin{equation}
                f_{A_i}(x, y, t) = \frac{G_i(x, y, t)}{\iint_R\int_{T_0}^{T_1} G_i(x, y, t) \,dt\,dx\,dy} = \frac{G_i(x, y, t)}{N_{A_i}}.
            \end{equation}
        
        \end{enumerate}

        We now construct a question-driven ensemble (QDE) model $E^{klm}$ as follows.
        The number questions (\ref{qu:nb}) and (\ref{qu:na}) are answered with model $M_k$,
        the background density question (\ref{qu:db}) is answered with model $M_l$,
        and the aftershock density question (\ref{qu:da}) is answered with model $M_m$.
        Note that questions (\ref{qu:nb}) and (\ref{qu:na}) are addressed with the same model.
        This is a choice made to avoid unrealistic event numbers.
        If one model interprets the majority of events as background, and another model interprets the majority of events to be aftershocks, answering the two questions with two different models would lead to exceptionally high or low total event numbers, which is not intended by the two ingredient models.
        
        In the notation above, which identifies a model with its parameters, this would give us
        
        \begin{equation}
            E^{klm} = (\mu_l \cdot \frac{N_{B_k}}{N_{B_l}}, \kappa_m \cdot \frac{N_{A_k}}{N_{A_m}}, c_m, \omega_m, \tau_m, d_m, \gamma_m, \rho_m).
        \end{equation}
        
\section*{Forecasting Experiments}
    \label{sec:forecast}
    To test whether flETAS models and QDE models which consist of ETAS and flETAS models provide better forecasts, we conduct pseudo-prospective forecasting experiments for Southern California and Italy.

    \subsection*{Competing Models}
        
        In these experiments, we consider the following four competing ingredient models.

        \begin{itemize}
            \item $M_0$: standard ETAS
            \item $M_1$: flETAS with free productivity and standard background
            \item $M_2$: flETAS with standard productivity and free background
            \item $M_3$: flETAS with free productivity and free background
        \end{itemize}   
        
        Out of these, $4^3=64$ QDE models can be constructed.

        Note that $M_2$ is conceptually close to the models described by \textcite{zhuang2012long} and \textcite{nandan2021seismicity}.
        
    \subsection*{Evaluation Metric}
        \label{sec:eval}

        We use interevent time horizons: Whenever an event occurs, a forecast is issued, which is valid until the occurrence of the next event.
        A pseudo-prospective model evaluation then aims to capture how well a forecast issued using data until event $e_{j-1}$ can describe the occurrence of the next event $e_j$.
        
        An ETAS forecast always consists of the forecasted background seismicity rate plus the forecasted aftershock seismicity rate.
        With this flexible definition of forecasting horizon, our ETAS forecast can be calculated and evaluated analytically.
        
        Consider $\lambda_i(t, x, y | \mathcal{H}_{t_{j-1}})$, the event rate under model $M_i$ as of time $t_{j-1}$ of the $(j-1)^{th}$ earthquake.
        This formulation of $\lambda_i$ is valid for times $t\in (t_{j-1}, t_{j}]$ between the occurrence of event $e_{j-1}$ and event $e_{j}$, and hence this is the forecasting horizon we consider.
        
        For the traditional experiment settings where one is interested in the seismicity forecast of the next days, months, or years, such an analytical description of the forecasted seismicity is not possible.
        As soon as an event occurs during the forecasting period, its aftershocks are not part of the background seismicity, nor of the aftershock seismicity that was calculated at the start of the forecasting period.
        For this reason, ETAS forecasts for fixed forecasting horizons are usually produced through the simulation of a large number of possible continuations of the catalog.

        In our case of flexible forecasting horizons, the log likelihood of observing $e_{j}$ under model $M_i$ is analytically defined (see \cite{ogata2013comprehensive}; \cite{daley2003introduction}) as
        
        \begin{equation}
            \ln \mathcal{L}_i (e_{j}) = \ln \lambda_i(t_{j}, x_{j}, t_{j} | \mathcal{H}_{t_{j-1}}) - \iint_R\int_{t_{j-1}}^{t_{j}} \lambda_i(t_{j}, x_{j}, t_{j} | \mathcal{H}_{t_{j-1}}) \,dt\,dx\,dy.
        \end{equation}
        
        We then define the information gain $IG^{i_1, i_2}_j$ of model $i_1$ over model $i_2$ during the $j^{th}$ forecasting period $(t_{j-1}, t_{j}]$ as
        
        \begin{equation}
            IG^{i_1, i_2}_j = \ln \frac{\mathcal{L}_{i_1}(e_j)}{\mathcal{L}_{i_2}(e_j)} = \ln \mathcal{L}_{i_1}(e_j) - \ln \mathcal{L}_{i_2}(e_j).
        \end{equation}

        The information gain per event (IGPE) over forecasting periods $j_1, \dots, j_K$ is defined as 
        
        \begin{equation}
            \frac{1}{K}\sum_{k=1, \dots, K} IG^{i_1, i_2}_{j_k},
        \end{equation}
        
        the average of IGs over those testing periods. 
        
        Compared to evaluation techniques based on the simulation of large numbers of possible catalog continuations such as in \textcite{nandan2019forecastingfull} and \textcite{mizrahi2021embracing}, which are encouraged by CSEP (see \cite{savran2022pycsep}), this approach allows us to compare models much faster, accelerating the development and testing process.
        To apply these models operationally, where forecasts are required for a fixed time horizon, simulations would still be required.
        This evaluation approach allows us to save time when developing and selecting the model to be used operationally, and is especially useful for evaluating a large suite of QDE models.
        
    \subsection*{Data}
        For Southern California, we consider the ANSS comprehensive earthquake catalog (ComCat), in the polygon given by the vertices in Table \ref{tab:polygon}.
        We consider earthquakes of magnitude $M\ge 2.0$ from January 1, 2010 until January 1, 2022.
        The first two years serve as auxiliary period in the ETAS and flETAS parameter inversion, and thus the start of the primary catalog is January 1, 2012.
        This means that the events between January 2010 and January 2012 can act as triggering events during the inversion, but not as triggered events.
        Using the method described by \textcite{mizrahi2021effect}, we find that the overall catalog is complete at this threshold, although there are likely periods during which the catalog is incomplete due to short-term aftershock incompleteness (STAI).
        Although \textcite{mizrahi2021embracing} have proposed a method to account for STAI in the ETAS model, we do not address this issue here.\\
        
        For Italy, we consider the Italian Seismological Instrumental and Parametric Data-Base catalog (ISIDe, \cite{iside2007italian}), in the area defined for the first CSEP experiment (\cite{schorlemmer2010setting}, vertices given in Table \ref{tab:it_polygon}).
        We consider earthquakes of magnitude $M\ge 2.5$ from April 16, 2005 until July 1, 2021.
        This is the time horizon available to modelers in the upcoming prospective CSEP forecasting experiment in Italy, and the estimated magnitude of completeness provided in the experiment description.
        The start of the primary catalog is January 1, 2010.
        
    \subsection*{Experiment Setting}
        For Southern California, we consider 5 years of testing, with the start of the first forecasting period at the occurrence of event $e_0$, the first event at or after January 1, 2017.
        In Italy, we consider 3 years of testing, starting at the occurrence of the first event at or after July 1, 2018.
        The idea of the pseudo-prospective experiments is to only use data that would have been available at the time the forecast is issued to calibrate the models.
        One could thus re-calibrate the model at the start of each forecasting period, whenever one more event becomes part of the catalog.
        To limit the number of computationally expensive parameter inversions for these experiments, we re-estimate the model parameters every 7 days in Southern California, and every day in Italy, and use the latest available set of parameters at the start time of each forecasting interval.
        Note that this does not mean that events between the calibration time and forecasting start are ignored.
        Their aftershocks are still considered in the calculated aftershock rate.
        We chose a shorter parameter updating interval for Italy to mimic the conditions of the CSEP experiment, and a longer one for Southern California to limit computational cost.
        
        We then calculate $IG^{i_1, i_2}_j$ for all $j$, and for all pairs of models $M_{i_1}, M_{i_2}$.
        If the IGPE over all forecasting periods of one model to another is positive, we consider the model to produce superior forecasts.
        \\
        
        As one could argue that generating a large number of models and then selecting the best performing ones somewhat invalidates the pseudo-prospective nature of our experiments, we consider the following additional model.
        At the start of the $j^{th}$ forecasting period, the total information gain of all QDE models during the last $n$ forecasting periods, i.e. periods $j-(n+1)$ to $j-1$, is compared. 
        The model with the highest IG is selected to produce the forecast for the $j^{th}$ forecasting period.
        We call this model QDE-S$_n$.
        
        This type of model, if capable of producing a powerful forecast, would be well suited to be used in an OEF context.

\section*{Results and discussion}
    \label{sec:results}
    The parameters that were obtained using the flETAS inversion algorithm are described in \nameref{sec:params} in the Appendix.
    Here, we present the results of the forecasting experiments.
 
    \subsection*{Experiment results}

\begin{figure}[H]
        \centering
        \includegraphics[width=\textwidth]{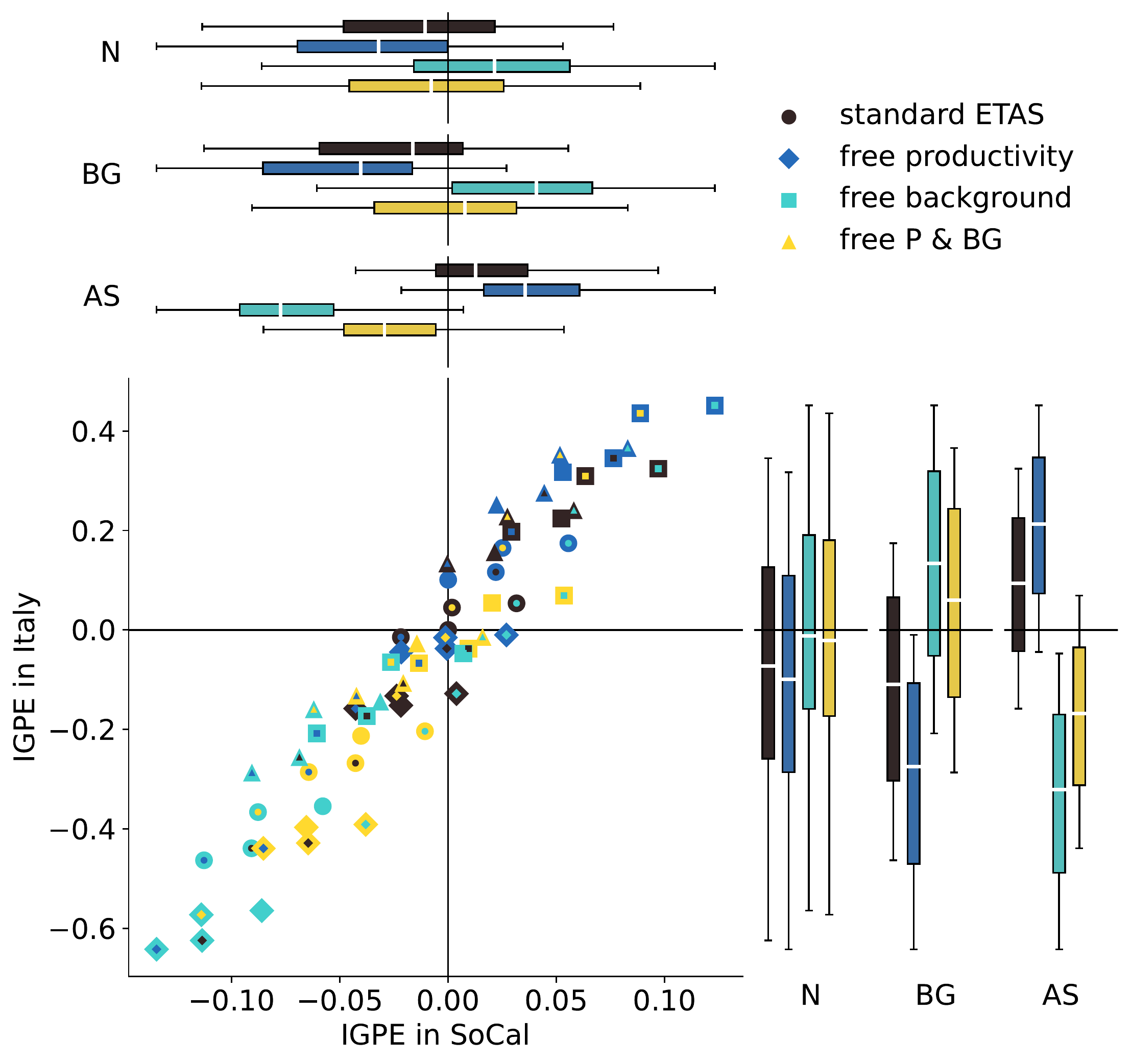}
        \caption{Scatter plot of IGPE over standard ETAS of the 64 QDE models in Italy and Southern California. Symbol shape, fill color and edge color describe the composition of the QDE. Shape, fill color, and edge color represent the ingredient model used to answer the background density (BG), number (N), and aftershock density (AS) questions, respectively. 
        Box plots on top (for Southern California) and to the right (for Italy) of the scatter plot: For N, BG, and AS questions, the four boxes represent the IGPE of four groups of QDE models. Each group contains the 16 QDE models which use a specific ingredient model (indicated by box color) to answer the indicated question.}
        \label{fig:ranks}
    \end{figure}

        Figure \ref{fig:ranks} compares the information gain per earthquake (IGPE) over the standard ETAS null model ($M_0 = E^{000}$) of all 64 QDE models in Italy and Southern California.
        The IGPE varies between -0.64 and 0.45 in Italy, and between -0.13 and 0.12 in Southern California.
        The best and worst performing QDE models are $E^{221}$ and $E^{112}$, respectively, for both regions.
        The best performing model $E^{221}$ uses the free background model $M_2$ to answer the number and background density questions, and the free productivity model $M_1$ to answer the aftershock density question.
        Vice versa, the worst performing model $E^{112}$ uses $M_1$ to answer the number and background density questions, and model $M_2$ to answer the aftershock density question. 
        Generally, the models which perform well or poorly in Italy are also performing similarly in Southern California.
        
        The symbol shape, fill color, and edge color in the scatter plot of Figure \ref{fig:ranks} represent the ingredient model used to answer the background density (BG), number (N), and aftershock density (AS) questions, respectively.
        Models which perform well tend to answer the BG question with the free background ingredient model, and the AS question with the free productivity model.
        Conversely, models which address the BG question with the free productivity model, and those which address the AS question with the free background model, tend to perform poorly.
        
        This is highlighted in the box plots of Figure \ref{fig:ranks}.
        There, for each question, the distribution of IGPE of the 64 QDE models is given per possible answer.
        While for the number questions, no clear trend can be inferred, it is evident that the free background model serves well at answering the BG question and the free productivity model serves well at answering the AS question.
        These trends are qualitatively very similar in Southern California and Italy.\\
        
        These results emphasize the added value generated by the flETAS approach, although most flETAS models individually do not outperform standard ETAS.
        Apparently, a model which gives full flexibility to the background rate during parameter inversion is more informative than others when addressing the background density question.
        And a model which is flexible at identifying aftershocks is more informative than others when answering the aftershock density question.
        These observations are made for both considered regions.
        
        While conceptually it makes sense that a model which can more flexibly capture one particular aspect of seismicity is particularly successful at answering questions about this very aspect of seismicity, this is simultaneously a somewhat counter-intuitive result.
        If flETAS with free background is more successful than other models at identifying background events, one would expect it, due to the self-consistent nature of parameter inversion, to also be more successful at identifying aftershocks, and thus at describing their occurrence times and locations.

        A possible interpretation of the observation that $E^{221}$, $E^{220}$, and even $E^{223}$ can so clearly outperform $E^{222}$, is the following.
        Compared to the null model $M_0$, model $M_2 = E^{222}$ allows the background seismicity to be free and therefore interprets a higher fraction of events in the training catalog to be background earthquakes, which manifests in a much higher background rate.
        $M_2$ can thus explain the spatial distribution of background events well, as well as the partitioning of seismicity into background events and aftershocks.
        Possibly, $M_2$ overestimates the background portion of the training catalog due to ``too much freedom".
        The level of overestimation may be small enough so that $M_2$ still captures the fraction and locations of background earthquakes better than the other ingredient models do.
        Overestimation of the background seismicity comes with underestimation of the fraction of aftershocks in the training catalog.
        While this underestimation may have a minor biasing effect on the number of background earthquakes and aftershocks, the spatio-temporal distribution of aftershocks can be affected in a more harmful way.
        Aftershocks which occur in the tails of the spatial or temporal distributions have higher chances to be falsely identified as background events compared to aftershocks which are close to their parent event. This leads to a distorted characterization of the aftershock triggering behavior of model $M_2$, which can be fixed by using the triggering parameters from models $M_0$ or $M_1$, as indicated by the good performance of models $E^{221}$ and $E^{220}$.\\

        Another noteworthy observation is that model $M_3$, which in principle has all the flexibility necessary to encompass the parameterization of model $E^{221}$, is clearly outperformed by $E^{221}$.
        We interpret this to be a consequence of the fact that the information which is optimized during model calibration and the information used for forecasting are not the same.
        This does not indicate a flaw in the method presented, but rather illustrates a complexity of the forecasting problem to which the QDE approach offers an apparently useful solution.

        

\begin{figure}[H]
        \centering
        \includegraphics[width=\textwidth]{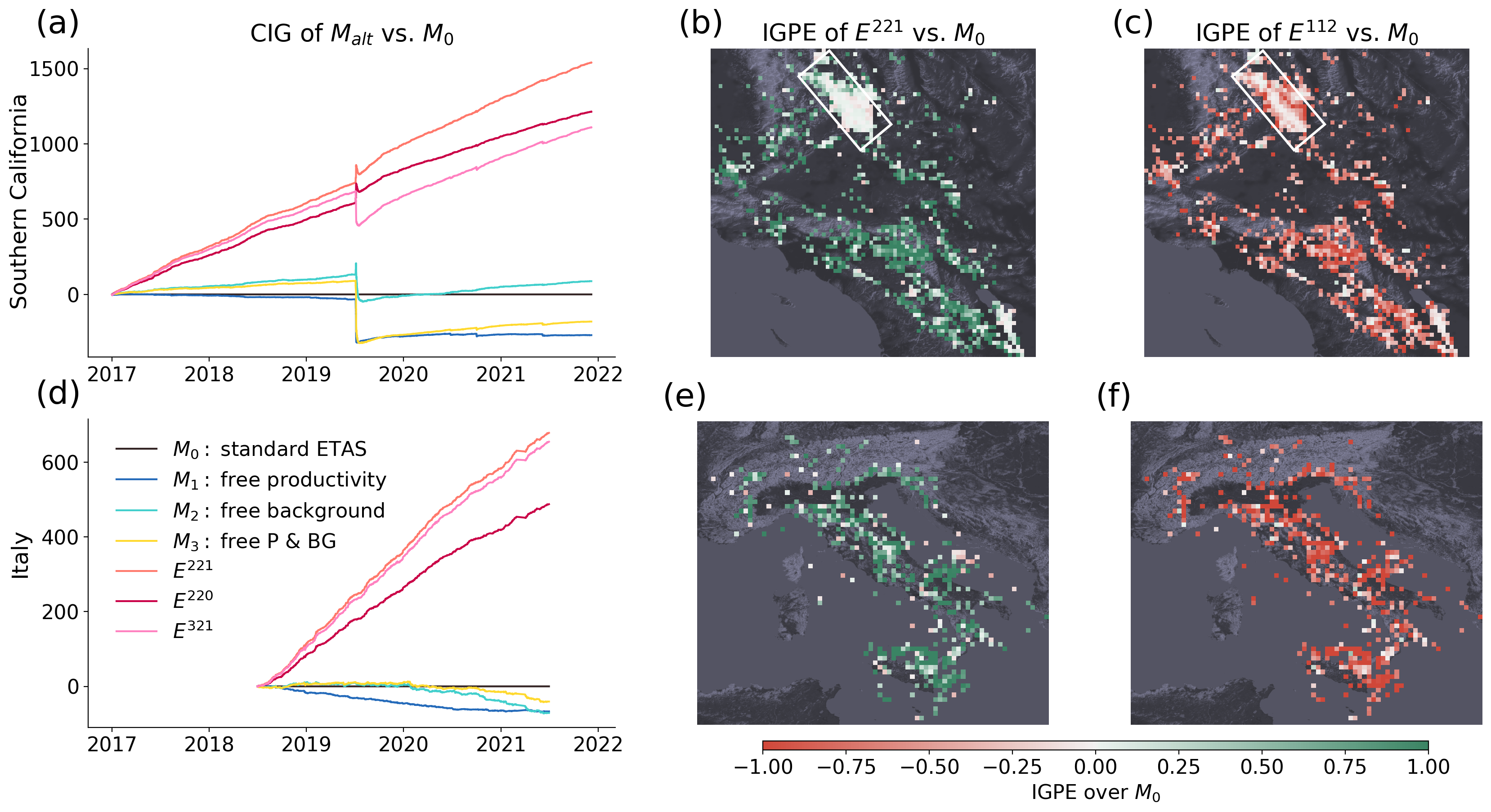}
        \caption{Top panels show results for Southern California, bottom panels for Italy.
        (a) and (d): Cumulative Information Gain (CIG) over time of the ingredient models and the three QDE models best performing in Southern California, compared to the standard ETAS model indicated by the black horizontal line.
        (b-c) and (e-f): Information gain per earthquake (IGPE) per spatial grid cell of the best performing QDE model ($E^{221}$, (b) and (e)) and the worst performing QDE model ($E^{112}$, (c) and (f)), compared to standard ETAS ($M_0 = E^{000}$).
        Grid cell resolution is 0.05 $\times$ 0.05 degrees in SoCal, and 0.2 $\times$ 0.2 degrees in Italy, chosen for best visibility.
        The white rectangle in (b-c) highlights the region of the Ridgecrest sequence in 2019.
        }
        \label{fig:CIG}
    \end{figure}

        Figure \ref{fig:CIG}(a) shows the cumulative information gain (CIG) over the standard ETAS model over time of the three flETAS ingredient models, and the three best performing QDE models.
        The CIG of model $i_1$ over model $i_2$ at time $t$ is given as the sum of IGs of all forecasting periods ending prior to time $t$,

        \begin{equation}
            \sum_{j: t_j < t} IG^{i_1, i_2}_j.
        \end{equation}
        
        In Southern California, the flETAS ingredient models have a negative information gain following the Ridgecrest events in July 2019, meaning that during this time, the standard ETAS model ($M_0$) is better performing.
        The free background model $M_2$ outperforms $M_0$ immediately after the onset of the sequence, and suffers from information loss later during the sequence.
        The other two ingredient models do not exhibit the initial information gain.
        Among the flETAS models, only $M_2$ can compensate for the information loss during the course of the 5 years of testing and ends up with a positive overall information gain.
        
        Among the QDE models presented, models $E^{221}$ and $E^{220}$ show an initial information gain after the onset of the Ridgecrest sequence, followed by a period of information loss.
        In contrast to the ingredient models, the information loss during the sequence is smaller than the gain at the beginning of the sequence, such that these models show positive information gain during the Ridgecrest sequence.
        The three QDE models in Figure \ref{fig:CIG}(a) also show a rapidly accumulating information gain throughout the testing period, arriving at an overall IGPE of 0.12, 0.10 and 0.09.
        
        From Figure \ref{fig:CIG}(b), it is clear that the IGPE is relatively close to zero in the Ridgecrest area, and the positive IG during the sequence must come from a few specific locations.
        In the rest of Southern California, higher IGPE values are achieved, with a median grid-cell-wise IGPE of 0.66 for model $E^{221}$ shown in \ref{fig:CIG}(b).
        Conversely, the median grid-cell-wise IGPE for the worst performing model $E^{112}$ shown in \ref{fig:CIG}(c) is -0.54.
        Generally, it performs poorly where $E^{221}$ performs well.
        \\

        In Italy, all flETAS models have negative total information gain over $M_0$.
        Nevertheless, two of the top three QDE models which perform best in Southern California are also among the top three in Italy, with overall IGPE values of 0.45 and 0.44 for $E^{221}$ and $E^{321}$.
        The second best model of SoCal, $E^{220}$, ranks sixth in Italy with an IGPE of 0.32.
        Similar to what can be observed in Southern California, the regions in Italy in which the best performing model $E^{221}$ performs well coincide with the areas in which model $E^{112}$ shown in Figure \ref{fig:CIG}(f) performs poorly.
        The median grid-cell-wise IGPE of the two models are 0.76 and -0.82, respectively.
        Although these grid-cell-wise IGPE values cannot directly be compared between Italy and Southern California due to the different size of the grid cells, the results suggest a qualitatively more similar model performance between the two regions than what is shown by the overall IGPE shown in Figure \ref{fig:ranks}.
        The lower IGPE in SoCal is likely caused by a relatively small IG during the Ridgecrest sequence when a large fraction of events occurred.

    \subsection*{Pseudo-prospective model selection}
        
\begin{figure}[H]
        \centering
        \includegraphics[width=\textwidth]{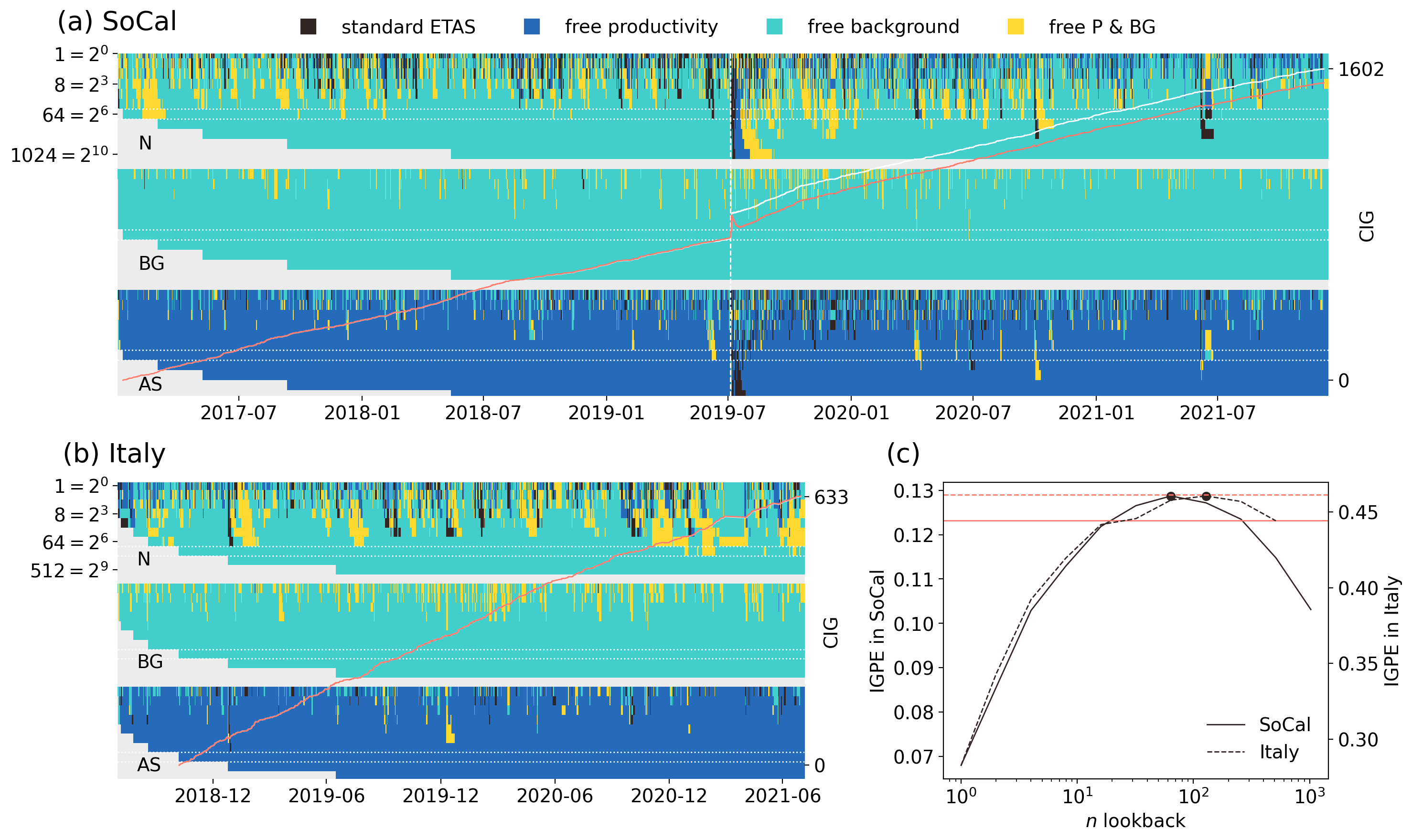}
        \caption{Composition and performance of QDE-S$_n$ models. (a) and (b) for Southern California and Italy: Composition of QDE-S$_n$, where $n$ takes values of powers of 2. Top, middle, and bottom part represent the ingredient model used to answer the number (N), background density (BG), and aftershock density (AS) questions. Within each part, $n$ increases from top to bottom. Dotted white lines highlight the best performing QDE-S$_n$. Solid white and orange line show the cumulative information gain (CIG) of the best QDE-S$_n$ and best QDE ($E^{221}$), respectively, for the period in which both are defined. 
        White line is barely visible for Italy because it coincides with the orange line.
        Vertical dashed line indicates the occurrence time of the M6.4 Ridgecrest event on July 04, 2019. 
        (c): IGPE of different QDE-S$_n$ (black lines), for different values of $n$. Horizontal orange lines indicate IGPE of $E^{221}$ for the period in which the best QDE-S$_n$ is defined. Solid lines represent Southern California, dashed lines represent Italy.
        }
        \label{fig:winner}
    \end{figure}

        Figure \ref{fig:winner} illustrates the composition and performance of QDE-S$_n$ models.
        The number $n$ of past forecasting periods considered when selecting the forecasting model for the next period is in $\{1=2^0, 2, 4, 8, 16, 32, 64, 128, 256, 512, 1024=2^{10}\}$ for SoCal, and $n\in\{1=2^0, \dots, 512=2^{9}\}$ for Italy.
        We do not consider $n=1024$ for Italy, as this would reduce the number of testing periods in which QDE-S$_n$ is defined by more than half compared to the QDE models.
        The top, middle, and bottom parts of Figure \ref{fig:winner}(a) and (b) show the ingredient model used by QDE-S$_n$ to answer the N, BG, and AS questions over time.
        Within each part, $n$ increases from top to bottom.
        As expected, the composition of QDE-S$_n$ is more stable as $n$ increases, and is almost always defined via $E^{221}$ for large $n$, in both regions.
        
        In Southern California, a change in composition can be observed after the onset of the Ridgecrest sequence in July 2019.
        Specifically, the number questions are best answered by standard ETAS, free productivity flETAS, and free productivity and background flETAS, in this order, before moving back to answering with free background flETAS.
        The aftershock question intermittently best answered by standard ETAS during the sequence.
        It is interesting to note here that the performance of $E^{221}$ and QDE-S$_{64}$ are almost identical throughout the 5 years of testing, with the difference that QDE-S$_{64}$ does not show the information loss after the initial information gain after the onset of the sequence.
        This results in an overall IGPE of 0.13 and 0.12 for QDE-S$_{64}$ and $E^{221}$, during the period in which both are defined, as is shown in Figure \ref{fig:winner}(c).
        Thus, the QDE-S$_n$ model, which was originally designed to avoid a biased selection of the winning model after knowing the experiment outcome, is capable of outperforming the winning QDE model for good choices of $n$, and clearly outperforms all ingredient flETAS models for any tested choice of $n$.
        
        In Italy, the best performing QDE-S$_n$ model is QDE-S$_{128}$.
        It is almost always using $E^{221}$ to issue a forecast for the next period, and thus unsurprisingly achieves the same IGPE.
        As in SoCal, all tested choices of $n$ yield a model which clearly outperforms all ingredient flETAS models.
        The most simple QDE-S$_n$ model, QDE-S$_1$, which always selects the best QDE model of the previous forecasting period to issue the next forecast, already achieves a very high IGPE of 0.28.\\

\section*{Conclusions}
    \label{sec:concl}
    
    We describe an adapted ETAS expectation maximization (EM) algorithm which allows a non-parametric inversion of aftershock productivity and/or background rate.
    Further, we introduce a novel approach of question-driven ensemble (QDE) modeling, which combines ingredient models by using them to answer different forecasting sub-problems.
    In pseudo-prospective forecasting experiments for Southern California and Italy, we compare the forecasting skill of three flexible ETAS (flETAS) models and a total of 60 nontrivial QDEs of flETAS and ETAS models, to that of the standard ETAS null model.
    
    We find that the best models tend to use flETAS with free background to model the number of events and locations of background earthquakes, and flETAS with free productivity to model the time and location of aftershocks.
    The best model is the same in both regions and achieves an information gain per earthquake (IGPE) over standard ETAS of 0.12 in Southern California, and 0.45 in Italy.\\
    
    To address the possible concern of a biased selection of the winning model after knowing the experiment outcome, we also test the forecasting skill of a model which pseudo-prospectively selects the currently best performing QDE model to issue the forecast for the next testing period.
    Depending on the criteria to identify the best QDE model, we find that the forecasting skill can be greater than that of the overall best QDE model.
    This approach thus provides a promising candidate for an operational earthquake forecast.

    During the 2019 Ridgecrest sequence in Southern California, different ingredient models are best suited to model the number of events during different stages of the sequence.
    The idea of operationally selecting different QDE models (i.e. selecting different ETAS model parameters) based on their recent performance is in this case related to the idea of \textcite{page2016three}.
    They considered sequence-specific parameters to be sampled from an underlying distribution and described a Bayesian approach to update this distribution as aftershock data becomes available.
    \\
    
    Our results can also be viewed as a first step toward developing a potentially fruitful branch of earthquake forecasting research.
    Several key questions remain open and are to be addressed in future studies:
    \textit{Why do QDE models outperform ingredient models which were inverted in a self-consistent way?
    What drives the success of different QDE models during different phases of the Ridgecrest sequence?
    How does QDE performance increase when further ingredient models are considered?
    And what does all of this teach us about the dynamics of seismicity?}

\section*{Data and Resources}
    The Advanced National Seismic System (ANSS) Comprehensive Earthquake Catalog (ComCat) provided by the U.S. Geological Survey (USGS) was searched using https://earthquake.usgs.gov/data/comcat/ (last accessed January 2022).
    The Italian Seismological Instrumental and Parametric Data-Base (ISIDe) was used as provided by the organizers of the upcoming CSEP experiment in Italy, and can be accessed via http://terremoti.ingv.it/en/search.

\section*{Acknowledgments}
   This study has been funded by the Eidgenössische Technische Hochschule (ETH) research grant for project number 2018-FE-213, “Enabling dynamic earthquake risk assessment (DynaRisk)”, the European Union's Horizon 2020 research and innovation program under Grant Agreement Number 821115, real-time earthquake risk reduction for a resilient Europe (RISE), the National Science Foundation (grant EAR-2122168), and the Southern California Earthquake Center (based on NSF Cooperative Agreement EAR-1600087 and USGS Cooperative Agreement G17AC00047). The paper benefited from constructive comments by the Associate Editor and two anonymous referees. 

    
    
    

\printbibliography[heading=bibintoc]

\section*{Full mailing address of each author}
    \begin{itemize}
        \item Leila Mizrahi\\leila.mizrahi@sed.ethz.ch\\Sonneggstrasse 5\\8092 Zürich\\Switzerland
        \item Shyam Nandan\\snandan@ethz.ch\\Sonneggstrasse 5\\8092 Zürich\\Switzerland
        \item William Savran\\wsavran@usc.edu\\University of Southern California\\90007 Los Angeles, CA\\United States
        \item Stefan Wiemer\\stefan.wiemer@sed.ethz.ch\\Sonneggstrasse 5\\8092 Zürich\\Switzerland
        \item Yehuda Ben-Zion\\benzion@usc.edu\\University of Southern California\\90007 Los Angeles, CA\\United States
    \end{itemize}

\renewcommand{\listfigurename}{List of Figure Captions}
\listoffigures
\section*{Figures}

\newpage
\let\svaddcontentsline\addcontentsline
\renewcommand\addcontentsline[3]{%
  \ifthenelse{\equal{#1}{lof}}{}%
  {\ifthenelse{\equal{#1}{lot}}{}{\svaddcontentsline{#1}{#2}{#3}}}}
\appendix
\setcounter{table}{0}
\setcounter{figure}{0}
\linespread{1.0}
\renewcommand{\thetable}{A\arabic{table}}
\renewcommand{\thefigure}{A\arabic{figure}}

    \section*{Appendix}
    \subsection*{Polygons}
    The polygons used in this study are defined via the following lists of vertices.

    \begin{table}[H]
    \begin{center}
            \caption{Southern California polygon boundary vertices.} 
            \begin{tabular}{lll}
            latitude & longitude &  \\\hline
            32.7219  & -116.3004 &  \\
            33.7424  & -117.6512 &  \\
            33.7958  & -117.966  &  \\
            33.9322  & -118.0775 &  \\
            34.0984  & -118.2611 &  \\
            34.1755  & -118.9365 &  \\
            34.6027  & -118.8775 &  \\
            34.8281  & -119.343  &  \\
            36.525   & -119.1988 &  \\
            36.4835  & -115.6381 &  \\
            34.128   & -115.5463 &  \\
            32.7219  & -115.2578 &  \\
            32.6922  & -115.448  &  \\
            32.7753  & -115.7234 &  \\
            32.8109  & -115.8545 & 
            \end{tabular}
            
            \label{tab:polygon}
    \end{center}

    \end{table}
    
    \begin{table}[H]
    \begin{center}
            \caption{Italy polygon boundary vertices.}
            \begin{tabular}{lll}
            latitude & longitude &  \\\hline
            45.1                 & 4.9                  \\
            44.5                 & 5.1                  \\
            43.3                 & 5.9                  \\
            42.8                 & 6.5                  \\
            41.6                 & 9.1                  \\
            38.0                 & 10.5                 \\
            36.7                 & 11.5                 \\
            35.8                 & 13.4                 \\
            35.3                 & 15.1                 \\
            35.7                 & 16.1                 \\
            38.8                 & 19.4                 \\
            40.1                 & 20.1                 \\
            41.3                 & 19.5                 \\
            42.9                 & 17.2                 \\
            44.0                 & 15.6                 \\
            45.6                 & 15.6                 \\
            46.5                 & 15.4                 \\
            47.5                 & 14.7                 \\
            47.9                 & 13.7                 \\
            48.1                 & 13.2                 \\
            48.4                 & 12.2                 \\
            48.2                 & 10.7                 \\
            47.9                 & 9.4                  \\
            47.8                 & 8.4                  \\
            46.8                 & 5.8                  \\
            45.8                 & 5.1                  \\
            45.1                 & 4.9                  \\
            \end{tabular}
             
            \label{tab:it_polygon}
    \end{center}

    \end{table}

    \newpage
    \subsection*{Inverted parameters}  
        \label{sec:params}
        
        \begin{figure}[H]
            \centering
            \includegraphics[width=\textwidth]{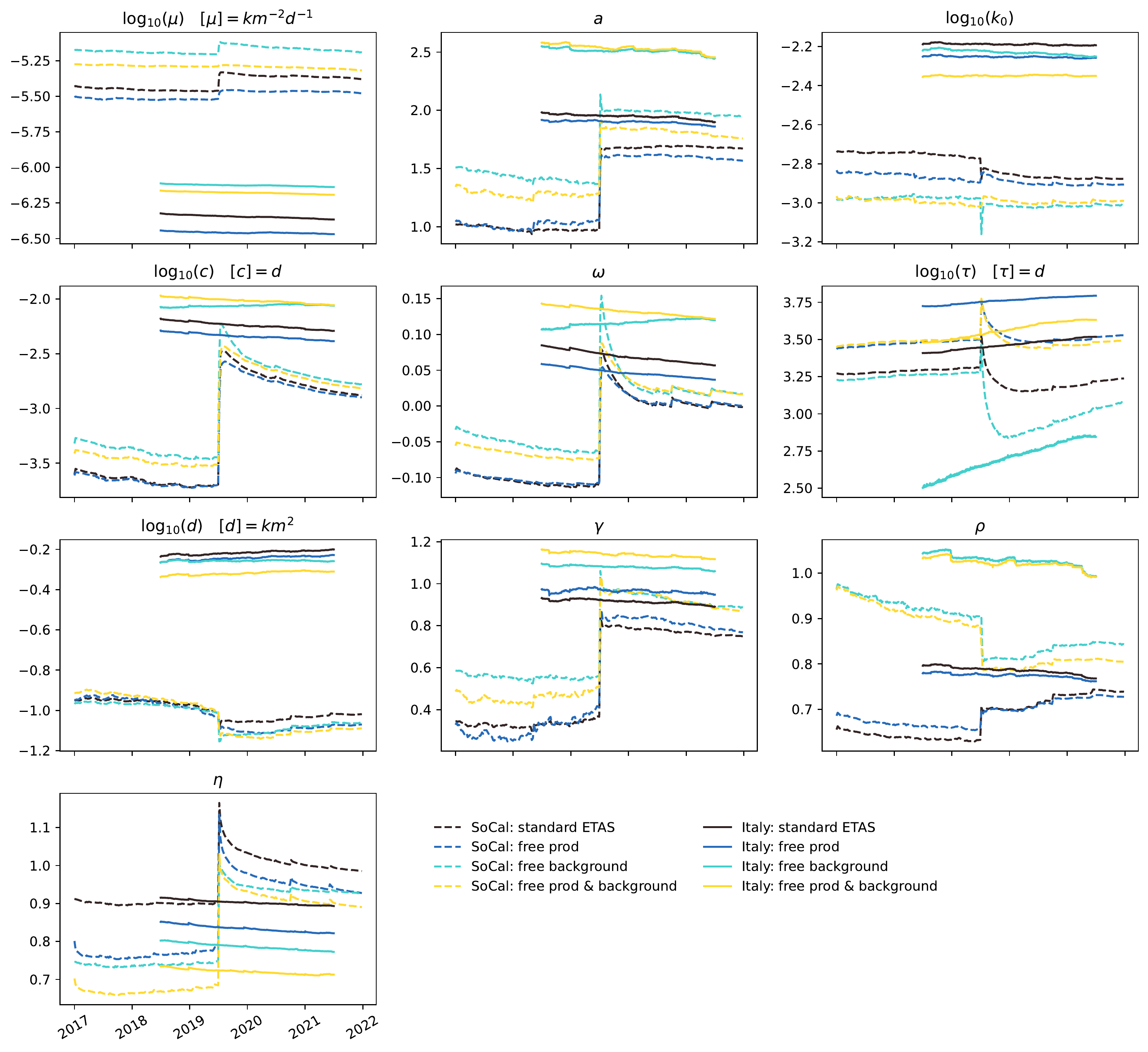}
            \caption{Evolution of inverted parameters with increasing length of the training catalog, for the four ingredient models. The branching ratio $\eta$ is not individually inverted, but is calculated from the other parameters. Dashed lines reflect Southern California parameters, solid lines reflect Italian parameters.}
            \label{fig:par_evol}
        \end{figure}
    
        Figure \ref{fig:par_evol} shows the inverted parameters for the four ingredient models, with an increasing time horizon used for the calibration, for Southern California and Italy.
        Note that for the standard ETAS model and flETAS where only the background rate is free, the parameters $a$ and $k_0$ are inverted directly during  expectation maximization (EM), while for the flETAS models with free productivity, they are inferred afterwards based on the $\kappa_j$ values that result from the EM inversion.\\

        Most parameters show remarkable changes in time in Southern California, and generally, the parameters differ between Italy and Southern California.
        The differences between parameters obtained for different ingredient models show similar trends in both regions.\\
        
        For instance, the background rate $\mu$ is highest for the model which only allows the background rate to be free, followed by the model where background and productivity are free, and is lowest when only the productivity is free.
        This is expected, since allowing the background to be free will allow the model to classify more events to be background events, while allowing the productivity to be free will allow it to classify more events to be aftershocks.
        
        The exponent of the productivity law, $a$, is larger in the flETAS models which allow the background to be free, indicating a stronger magnitude dependency of the number of aftershocks en earthquake is expected to generate.
        Those models also have larger $\gamma$ and much larger $\rho$ values, which translates to a stronger magnitude dependency of the spatial region in which aftershocks occur, and a stronger spatial decay of the aftershock rate.
        
        Interestingly, the flETAS model in which only productivity is free shows smaller $k_0$ values than standard ETAS in both regions, accompanied by values of $a$ that are similar to standard ETAS.
        Both these effects would suggest lower overall productivity.
        However, the value of $\tau$ is larger in this model, indicating a slower long-term tapering off of aftershock rate in time, and $\omega$ is smaller in Southern California (similar in Italy), further indicating a slower (similar) temporal decay of aftershock rate.
        Together with the observation that $\mu$ is smaller for this model, these results suggest that allowing productivity to free leads to an overall slower decay of aftershock rate, and thus a large fraction of aftershocks is expected to occur later in an ongoing sequence.
        
        The branching ratio $\eta$, which captures the average expected number of aftershocks of any event, is highest for the standard ETAS model, followed by flETAS with free productivity, flETAS with free background, and flETAS with free productity and background with the lowest branching ratio.
        Thus, the degree of flexibility of a model is qualitatively opposite to the degree of criticality of the system that is inferred with that model.
       
\end{document}